# THE XMM-NEWTON HARD BAND WIDE ANGLE SURVEY

Carangelo N. [1], Molendi S.[1] and the Hellas2XMM collaboration

[1]*Istituto di Astrofisica Spaziale e Fisica Cosmica (IASF)*
*Sez. di Milano - CNR, Via Bassini, 15, I-20133 Milano, Italy*

## 1. Introduction

The origin of the hard XRB as a superposition of unabsorbed and absorbed AGNs is now widely accepted as the standard model. It has been recognized that a self consistent AGN model for the XRB (e.g. Comastri et al. 1995; Gilli et al. 2001) requires a combined fit of several observational constraints in addition to the XRB spectral intensity, such as the number counts, the redshift and the average spectra. Furthermore, a key ingredient is the absorption distribution of the hydrogen column density $N_H$: this is a very critical parameter in the model, the only one not constrained by any direct measurement and treated as a free parameter in the fitting procedure. To date, synthesis models have been based on the local $N_H$ distribution of Piccinotti et al. (1982) sample or more recently on the Risaliti et al. (1999) distribution for Seyferts 2, even if several works have tackled the issue of $N_H$ distribution (Maiolino et al. 1998, Bassani et al. 1999, Bauer et al. 2003). However all of these kinds of study are still affected by selections effects or incompleteness problems or by biases related to the absorption.

We designed an XMM-Newton serendipitous survey in the hard band [7-11] keV in order to improve the Piccinotti et al. (1982) distribution and to obtain a more reliable $N_H$ distribution according to criteria independent of absorption effects. The selection of the hard [7-11] keV band can allow us to be unbiased against absorbed objects as much as possible. Moreover, the high throughput of XMM combined with the selection of bright objects, can allow us to perform an accurate spectral analysis and characterization of the sources.

The goal was to build a local reference sample the least affected than any other by $N_H$ bias in order to study its properties and compare them to those of other deeper and non local samples; to derive, from spectral modeling, a solid determination of $N_H$ followed by the $N_H$ distribution of our sample (the $N_H$ distribution has been usually derived from hardness ratio measurements in the previous surveys performed in the hard band); to help constrain the X-ray Luminosity Function (XRLF) of absorbed objects and consequently the AGN unification schemes and the synthesis models of the XRB; finally to address other important issues such as the relative importance of the reflection component and the equivalent width (EW) of the iron lines.

## 2. The sample

We analyzed all of the XMM-EPIC fields publicly available before March 1st 2003, selected at high galactic latitude $|b| > 27$ (excluding the Magellanic Clouds), with $t_{exp} > 10$ ksec and with a full frame observation in PN+MOS1+MOS2. Our survey covers about 30 square degrees for a total of 112 fields observed and 45 sources detected in the [7-11] keV band down to a limiting flux of about $3 \times 10^{-14}$ erg cm$^{-2}$ s$^{-1}$. We cross-correlated the X-ray positions of the 45 sources with the Nasa Extragalactic Database (NED), finding the optical counterpart for 10 objects with a measured redshift. For 16 sources we found an optical counterpart visible in the R plates (which have a limiting magnitude of 21) of the Digitized Sky Survey (DSS) and as far as the other sources with no DSS counterpart are concerned, we estimated the limiting optical magnitude from the Fiore et al. (2003) diagram (see figure 1). Given the high limiting flux of our survey ($F_{lim} \sim 5 \times 10^{-14}$ in the [2-10] keV band), we expected the remainder of our sources to have an optical magnitude of R< 22-23 and therefore, to be able to perform an optical follow-up with a 4m class telescope.

## 3. Data reduction and spectral characterization

The data were processed using the XMM-Science Analysis Software (XMM-SAS v5.4) and the Baldi et al. (2002) pipeline implemented for the [7-11] keV band. The excellent astrometry between the three cameras allowed us to merge together the MOS and PN images in order to increase the S/N of the sources.

We constructed an automatic spectral pipeline to extract, from the cleaned event files, the source spectrum, within a radius $r_{sou}$=30-45″ and the corresponding background spectrum within a radius $r_{bkg}$=3×$r_{sou}$. We obtained spectra for the PN, for the two MOS cameras and for the combined MOS1 and MOS2 cameras. With the final products we performed spectral characterization in the 0.4-10 keV energy range.

We carried out a preliminary spectral characterization in the [2-10] keV band, using a simple absorbed power law model in order to achieve a preliminary $N_H$ distribution: about 30-35% of our sources have a value of $N_H > 10^{22}$. We report the our preliminary average value in the figure of Baldi et al. (2003) (see figure 2): a clear inconsistency is present between the fraction derived from both the HR and spectral analysis and the prediction of the XRB synthe-sis model. We expected that our survey can allow us to put a more reliable constraint on this fraction by a direct spectral characterization in the flux range $5 \times 10^{-14}$-$5 \times 10^{-13}$ erg cm$^{-2}$ s$^{-1}$ in the 2-10 keV band and to test whether this inconsistency is true or related to bias effect.

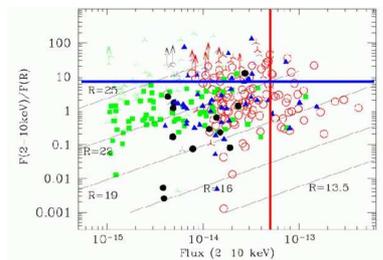 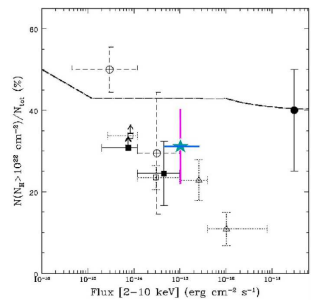

*Figure 1.* The X-ray ([2-10] keV) to optical (R band) flux ratio as a function of the X-ray flux combined sample:open circles= HELLAS2XMM, filled squares=CDFN, filled triangles=LH, filled circles=SSA13 (see Fiore et al. 2003 for details).Solid lines mark loci of constant R band magnitude. The vertical solid line mark the limiting [2-10] keV band flux of our survey.

*Figure 2.* The fraction of sources with $N_H > 10^{22}$ (at z=0) as a function of 2-10 keV flux (see Baldi et al. 2003 for details). Filled squares=HELLAS2XMM, void triangles= ASCA MSS, void circles =CDFS, filled circle= value of Piccinotti et al. (1982), filled star=our preliminary average. The thick dash-dotted line represents the prediction of the XRB synthesis model (Comastri et al. 2001)

## 4. A peculiar source

The very wide angle covered by our XMM-Survey allows us to search for very rare objects, like type 2 Quasars (QSO2): only a handful of QSO2 are known to date. Here we present the example of a very peculiar and interesting object found during our serendipitous survey. The complex spectrum of this source has been fitted in the 0.3-10 keV range with a *Model* described with the following formula:

$$Model = A_G \times (M_C + ZA_T \times (PL + R_C + GL))$$

where $A_G$ is the absorption associated with the Galactic column, $M_C$ is the thermal component,which describes the soft excess (MEKAL in XSPEC); $ZA_T$ is the absorption acting on the nuclear emission; PL is the power-law modeling the nuclear component with $\Gamma=1.8$; $R_C$ is the cold reflection component (PEXRAV in XSPEC); GL is a Gaussian line modeling the iron line at 6.4 keV. In figure 3 we report the PN spectrum and the best fitting model and residuals. It is clear that only the primary component is relevant. The iron line, detected at 97% significativity allowed us to measure the redshift of the source and the best fit value is z=0.31±0.03. We obtained a value of $N_H \simeq 10^{23}$ and at this redshift a luminosity in the 2-10 keV band of $L_{[2-10]}=1.4 \times 10^{44}$ erg cm$^{-2}$ s$^{-1}$. We conclude that this object is a QSO2. In particular, this is the first QSO2 discovered and completely spectral characterized in the X-ray (typically QSO2 were discovered with combined X-ray and optical observations).

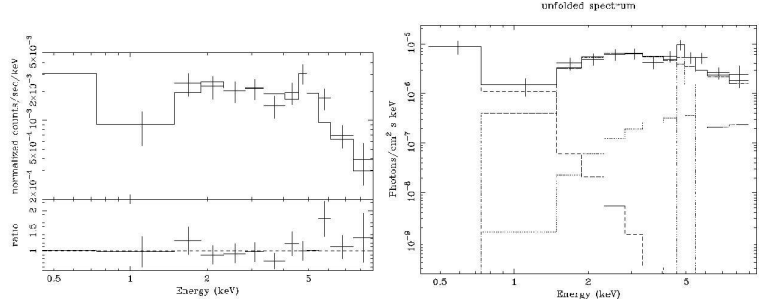

*Figure 3.* **Left Pannel**: The best fitting model and residual of QSO2. **Right Pannel**: The unfolded spectrum of QSO2. The dashed line is the thermal component, the dot-dashed line is the power law component, the dotted line is the reflection component and the dot-dot-dashed line is the iron line at 6.4 keV.

## 5. Conclusion

We would like to stress that our XMM-Survey performed in the hard [7-11] keV band allows us to perform an accurate spectral analysis and characterization for the first time of a local reference sample, the least affected than any other by $N_H$ bias in order to derive a $N_H$ distribution and constrain the XRLF of absorbed objects. Considering the knowledge and scientific equipment available today, this is the best result we could achieve. Future study will cover the optical follow up and we plan to extend our survey to about 40 square degrees before October 1st 2003.